\documentclass[twocolumn,english,aps]{revtex4}
\usepackage[T1]{fontenc}
\usepackage[latin9]{inputenc}
\usepackage{amsmath}
\usepackage{graphicx}
\usepackage{amssymb}
\usepackage{esint}

\usepackage{babel}

\begin{document}

\title{Theory of Cherenkov radiation in periodic dielectric media: Emission
spectrum}

\author{Christian Kremers, Dmitry N. Chigrin, and Johann Kroha}

\affiliation{Physikalisches Institut, Universit\"at Bonn, Nussallee 12, D-53115
Bonn, Germany}
\begin{abstract}
The Cherenkov radiation is substantially modified in the presence
of a medium with a nontrivial dispersion relation. We consider Cherenkov
emission spectra of a point or line charge, respectively, 
moving in general, three- (3D) and
two-dimensional (2D) photonic crystals. Exact analytical expressions
for the spectral distribution of the radiated power are obtained in
terms of the Bloch mode expansion. The resulting expression reduces
to a simple contour integral (3D case) or a one-dimensional sum (2D
case) over a small fraction of the reciprocal space, which is defined
by the generalized Cherenkov condition. We apply our method to a specific
case of a line source moving with different velocities in a 2D square-lattice
photonic crystal. Our method demonstrates a reasonable agreement with
numerically rigorous finite-difference time-domain calculations while
being less demanding on computational resources.
\end{abstract}
\maketitle
79.20.2m, 42.70.Qs, 41.60.Bq

\section{Introduction}

Back in 1934 Cherenkov reported the observation of the electromagnetic
radiation produced by an electron moving in a dielectric medium at
a velocity greater than the phase velocity of light in this medium
\cite{che34}. Such a radiation possesses a unique angular and frequency
spectrum and is called \emph{Cherenkov radiation} \cite{jel58}.
A nontrivial dispersion relation of a medium leads to substantial
modifications of the Cherenkov radiation. It has been shown that an
electron moving in a homogeneous medium with dispersion should emit
at any velocity \cite{fer40}. Richer spatial distribution of the
emitted radiation including intensity oscillations behind the Cherenkov
cone is a signature of the radiation in such a medium \cite{akm99,carb01,acrb03}.

To understand the properties of the Cherenkov radiation one can represent
the moving electron with space-time dependence of the corresponding
current density $\mathbf{J}\left(\mathbf{r},t\right)\sim\delta^3\left(\mathbf{r}-\mathbf{v}t\right)$
as a superposition of plane waves $\delta^3\left(\mathbf{r}-\mathbf{v}t\right)=\sum_{\mathbf{k}}\exp\left(i\mathbf{k}\cdot\mathbf{r}-i\mathbf{k}\cdot\mathbf{v}t\right)$
with different wave vectors $\mathbf{k}$ and frequency $\mathbf{k}\cdot\mathbf{v}$,
where $\mathbf{v}$ is the electron velocity. Only plane waves with
frequency and wave vector fitting the medium dispersion $\omega\left(\mathbf{k}\right)$
can resonantly excite electromagnetic modes in the medium, which gives
the Cherenkov resonance condition \cite{ft37}: \begin{equation}
\omega\left(\mathbf{k}\right)=\mathbf{k}\cdot\mathbf{v}.\label{eq:cherenkov-condition}\end{equation}
In a homogeneous, non-dispersive medium with refractive index $n$,
the dispersion relation is simply given by $\omega\left(\mathbf{k}\right)=\left(c/n\right)\left|\mathbf{k}\right|$
and the Cherenkov condition (\ref{eq:cherenkov-condition}) leads
to a well known conical wave front with an aperture $\cos\phi=c/\left(n\left|\mathbf{v}\right|\right)$
and a condition on the electron velocity $\left|\mathbf{v}\right|>c/n$
\cite{ft37,jel58}, $c$ being the vacuum speed of light. In an inhomogeneous
medium the interplay between interference and propagation can result
in an engineered nontrivial dispersion relation $\omega\left(\mathbf{k}\right)$.
For example, periodic dielectric media (photonic crystals) \cite{jmw95,sak01}
substantially modify both dispersion and diffraction of electromagnetic
waves possessing many unusual and novel optical phenomena, including
modification of emission dynamics \cite{byk72,yab87,jw90}, ultra-refraction
\cite{rus86,zen87,kos98,kos99} and photon focusing \cite{ep96,cst01,chi04}
effects. The present work focuses on an analytical understanding of the influence of a periodic medium
on the Cherenkov effect. 

\begin{figure}[b]
\begin{centering}
\includegraphics[width=1\columnwidth]{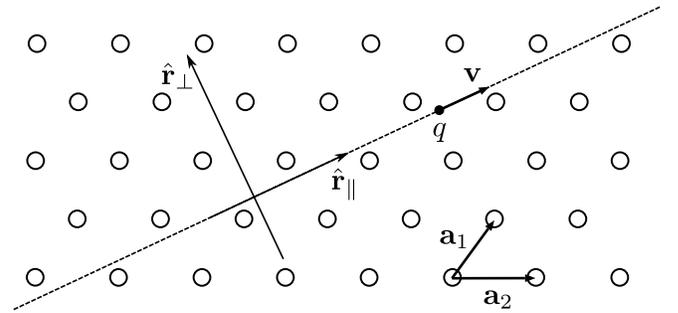}
\par\end{centering}

\caption{A sketch of a periodic medium and a charge trajectory. Basis
vectors $\mathbf{a}_{i}$ of the lattice are shown. The coordinate
system is chosen with one axis along ($\mathbf{\hat{r}}_{\parallel}$)
and the other perpendicular ($\mathbf{\hat{r}}_{\perp}$) to the charge
trajectory. \label{fig:structure_sketch}}

\end{figure}

Several studies on the modification of the radiation produced by a
charged particle moving near or inside periodic dielectric media are available.
The Cherenkov radiation in cholesteric liquid crystals has been analysed in Ref.~\cite{bel86}.
The modification of the Smith-Purcell radiation has been recently
studied both theoretically and experimentally near a surface of a
two- (2D) and three-dimensional (3D) photonic crystal \cite{abajo03b,oo04a,oo04b,yam04,oo06}.
The Cherenkov radiation generated by an electron moving inside an
air pore of a 2D photonic crystal perpendicular to the periodicity
plane has been used to map its photonic band structure in Refs.~\cite{abajo03,abajo03a}.
In all above mentioned reports, the theoretical analysis of the Cherenkov
effect has been done in the plane wave basis. Spatial and spectral
modifications of the Cherenkov radiation produced by an electron moving
in the periodicity plane of a 2D photonic crystal have been studied
in Ref.~\cite{lijj03} using the finite-difference time-domain (FDTD)
method. To date, there do not exist any reports on the general theory
of the Cherenkov effect in an arbitrary 3D periodic dielectric medium~\footnote{
From the fundamental point of view the parametric X-ray radiation from the moving electron in
crystrals has very similar physical nature and radiation mechanism, e.g. Ref.~\cite{fer05}.}.

The main purpose of the present work is to develop such a theory and
to provide a simple expression for the Cherenkov emission spectrum
(energy loss spectrum) for a point or line charge, respectively, moving with  
velocity $\mathbf{v}$
in an arbitrary direction inside a general, 3D or 2D photonic crystal.
To achieve this goal we derive an analytical expression for the power
emitted per unit length of the charge trajectory in terms of the Bloch
mode expansion. This expression is further reduced to a simple contour
integral (3D case) or a one-dimensional sum (2D case) over a small
fraction of the reciprocal space. As a result, to calculate the Cherenkov
emission spectrum, Bloch eigenmodes and their corresponding group
velocities are required only along an integration path (3D case) or
at a discrete set of $k$-points (2D case), considerably reducing
computational demands. The integration path and the discrete set of
points are defined by the generalized Cherenkov condition. Our theory
confirms that the Cherenkov radiation does exist in a periodic medium
for an arbitrary electron velocity \cite{lijj03}. It also predicts
an enhancement of the radiated power near the frequencies corresponding
to the vanishing component of the group velocity, which is orthogonal
to the electron trajectory.

The paper is organized as follows. In Section \ref{sec:Radiated-field}
the general solution of Maxwell's equations is summarized for an arbitrary
periodic medium. In Section \ref{sec:Emission-spectrum} an analytical
expression is derived for the power radiated per unit length by a moving point 
charge in 3D and a line charge in 2D periodic media. In Section \ref{sec:Numerical-example}
we apply our theory to calculate the Cherenkov emission spectra
in the particular case of a 2D photonic crystal. The predictions of the
analytical theory are substantiated by numerically rigorous FDTD calculations.
Section \ref{sec:Conclusions} concludes the paper.

\section{Radiated field\label{sec:Radiated-field}}

We consider a point (line) charge $q$  uniformly moving with a
velocity $\mathbf{v}$ along some direction in a general, infinite
periodic 3D or 2D medium $\varepsilon\left(\mathbf{r}\right)=\varepsilon\left(\mathbf{\mathbf{r}}+\mathbf{R}\right)$
(Fig. \ref{fig:structure_sketch}). Here $\mathbf{R}$ is a vector
of the direct Bravais lattice, $\mathbf{R}=\sum_{i}l_{i}\mathbf{a}_{i}$,
$l_{i}$ is an integer and $\mathbf{a}_{i}$ is a basis vector of
the periodic lattice. It is assumed that the medium is linear, nonmagnetic
and that no absorption takes place. Then the relevant Maxwell's equations read
in SI units:\begin{eqnarray}
\nabla\times\mathbf{E}\left(\mathbf{r},t\right) & = & -\mu_{0}\frac{\partial}{\partial t}\mathbf{H}\left(\mathbf{r},t\right),\label{eq:maxwell_rotE}\\
\nabla\times\mathbf{H}\left(\mathbf{r},t\right) & = & \varepsilon_{0}\varepsilon\left(\mathbf{r}\right)\frac{\partial}{\partial t}\mathbf{E}\left(\mathbf{r},t\right)+\mathbf{J}\left(\mathbf{r},t\right),\label{eq:maxwell_rotH}\end{eqnarray}
where, the electric (magnetic) field is denoted by $\mathbf{E}$ ($\mathbf{H}$).
An electromagnetic field is produced
by a current source $\mathbf{J}\left(\mathbf{r},t\right)$, which
in the case of the moving point (line) charge is defined as
\begin{equation}
\mathbf{J}\left(\mathbf{r},t\right)=q\mathbf{v}\delta^d\left(\mathbf{r}-\mathbf{v}t\right).\label{eq:current_time}
\end{equation}
Here $d=2,\,3$ is the dimensionality of the periodic
lattice. In the frequency domain, a general solution of Maxwell's equations (\ref{eq:maxwell_rotE}-\ref{eq:maxwell_rotH})
for an arbitrary current source $\mathbf{J}\left(\mathbf{r},t\right)$
and a periodic dielectric function $\varepsilon\left(\mathbf{r}\right)$
is given in terms of the Bloch eigenmode expansion \cite{dow92,sak01}\begin{widetext}
\begin{equation}
\mathbf{E}\left(\mathbf{r},\omega\right)=-\mathrm{i}\frac{1}{\left(2\pi\right)^{d}}\frac{\omega}{\varepsilon_{0}}\sum_{n}\int_{BZ}d^{d}k\int d^{d}r'\left\{ \frac{\mathbf{E}_{\mathbf{k}n}^{(T)}\left(\mathbf{r}\right)\otimes\mathbf{E}_{\mathbf{k}n}^{(T)\star}\left(\mathbf{r}'\right)}{\left(\omega-\omega_{\mathbf{k}n}^{(T)}+\mathrm{i}\gamma\right)\left(\omega+\omega_{\mathbf{k}n}^{(T)}+\mathrm{i}\gamma\right)}+\frac{\mathbf{E}_{\mathbf{k}n}^{(L)}\left(\mathbf{r}\right)\otimes\mathbf{E}_{\mathbf{k}n}^{(L)\star}\left(\mathbf{r}'\right)}{\left(\omega+\mathrm{i}\gamma\right)^{2}}\right\} \cdot\mathbf{J}\left(\mathbf{r}',\omega\right).\label{eq:E(r,w)_general}\end{equation}
\end{widetext}Here $\mathbf{J}\left(\mathbf{r},\omega\right)$ is the Fourier
transform of the current density $\mathbf{J}\left(\mathbf{r},t\right)$,
for the moving point (line) charge (\ref{eq:current_time}) given by %
\footnote{We use the following definition of direct and inverse Fourier transform:
$f\left(t\right)=\frac{1}{2\pi}\int_{-\infty}^{\infty}d\omega\, f\left(\omega\right)\mathrm{e}^{-\mathrm{i}\omega t}$
and $f\left(\omega\right)=\int_{-\infty}^{\infty}dt\, f\left(t\right)\mathrm{e}^{\mathrm{i}\omega t}$%
} \begin{equation}
\mathbf{J}(\mathbf{r},\omega)=q\hat{\mathbf{r}}_{\parallel}\delta^d(\mathbf{r}_{\perp})\exp\left(\mathrm{i}\omega\frac{r_{\parallel}}{\left|\mathbf{v}\right|}\right).\label{eq:current_omega}\end{equation}
The coordinate system is chosen with one axis, $\hat{\mathbf{r}}_{\parallel}$,
being parallel and other, $\left\{ \hat{\mathbf{r}}_{\perp}^{i}\right\} $,
orthogonal to the electron trajectory (Fig. \ref{fig:structure_sketch}).
$\mathbf{E}_{\mathbf{k}n}^{(T)}\left(\mathbf{r}\right)$ and $\mathbf{E}_{\mathbf{k}n}^{(L)}\left(\mathbf{r}\right)$
are generalized transverse and longitudinal Bloch eigenmodes \cite{dow92,sak01}
characterized by the band index $n$, the wave vector $\mathbf{k}$
and the eigenfrequencies $\omega_{\mathbf{k}n}^{(T)}$ and $\omega_{\mathbf{k}n}^{(L)}$,
respectively. The Bloch eigenmodes satisfy standard lattice periodic
boundary conditions. As it will be shown in the next Section, only
transverse Bloch eigenmodes contribute to the Cherenkov radiation field.
The asterisk ($\star$) and $\otimes$ denote the complex conjugate
and the outer tensor product in 3D space, respectively. Bloch eigenmodes
satisfy the homogeneous wave equation and fulfill the normalization
conditions\begin{widetext}\begin{equation}
\int d^{d}r\,\varepsilon\left(\mathbf{r}\right)\mathbf{E}_{\mathbf{k}n}^{(\alpha)\star}\left(\mathbf{r}\right)\cdot\mathbf{E}_{\mathbf{k}'n'}^{(\beta)}\left(\mathbf{r}\right)=\left(2\pi\right)^{d}\delta_{\alpha\beta}\delta_{nn'}\delta^d\left(\mathbf{k}-\mathbf{k}'\right)\label{eq:normalization}\end{equation}
and completeness relations\begin{equation}
\sum_{n\alpha}\int_{BZ}d^{d}k\,\sqrt{\varepsilon\left(\mathbf{r}\right)\varepsilon\left(\mathbf{r}'\right)}\mathbf{E}_{\mathbf{k}n}^{(\alpha)}\left(\mathbf{r}\right)\otimes\mathbf{E}_{\mathbf{k}n}^{(\alpha)\star}\left(\mathbf{r}'\right)=\left(2\pi\right)^{d}\hat{\mathbf{1}}\delta^d\left(\mathbf{r}-\mathbf{r}'\right),\label{eq:completeness}\end{equation}
\end{widetext}where $\alpha,\,\beta=T$ or $L$, and $\hat{\mathbf{1}}$
is the unit tensor. In Eqs. (\ref{eq:E(r,w)_general}) the $k$-space
integration is performed over the first Brillouin zone (BZ) of the
periodic medium and the summation is carried out over different photonic
bands. A positive infinitesimal $\gamma$ in (\ref{eq:E(r,w)_general})
assures causality \cite{sak01}.

\section{Emission spectrum\label{sec:Emission-spectrum}}

The emitted power of the Cherenkov radiation in a general periodic
medium is given by the rate at which the moving charge does work on
the surrounding electromagnetic field. For an arbitrary current density
$\mathbf{J}\left(\mathbf{r},t\right)$ in a 3D or 2D volume $V_{0}$,
the time-dependent emitted power is given by \cite{jac75} \begin{equation}
P\left(t\right)=-\intop_{V_{0}}d^{d}r\,\mathbf{J}\left(\mathbf{r},t\right)\cdot\mathbf{E}\left(\mathbf{r},t\right).\label{eq:power_time}\end{equation}
The total energy $U$ radiated by the current $\mathbf{J}\left(\mathbf{r},t\right)$
is obtained by integrating (\ref{eq:power_time}) over all moments
of time\begin{equation}
U=\int_{-\infty}^{\infty}dt\, P\left(t\right).\label{eq:total_energy}\end{equation}
The time integral in (\ref{eq:total_energy}) can be further transformed
into the integral over frequency (see Appendix A)\begin{equation}
U=\int_{0}^{\infty}d\omega\, P\left(\omega\right),\label{eq:total_energy_omega}\end{equation}
with a total power radiated per frequency interval $\left[\omega,\,\omega+d\omega\right]$
given by\begin{equation}
P\left(\omega\right)=-\frac{1}{\pi}\mathrm{Re}\left[\intop_{V_{0}}d^{d}r\,\mathbf{J}\left(\mathbf{r},\omega\right)\cdot\mathbf{E^{\star}}\left(\mathbf{r},\omega\right)\right].\label{eq:power_omega}\end{equation}
To obtain the power emitted per unit length of the charge trajectory
the integration volume $V_{0}$ should be chosen as a cylinder coaxial
with the trajectory, while the integral itself should be
normalized by the cylinder length $l$. In the 2D case, the volume
integral is reduced to the surface integral over a rectangle coaxial
with the charge trajectory and the result should be normalized to
the rectangle length.

We further derive the spectral dependence of the power $\left(dP/dl\right)$
(\ref{eq:power_omega}) radiated per unit length by the point (line) charge
(\ref{eq:current_time}) uniformly moving in a periodic medium. Assuming
that the presence of the moving charge does not change the band structure
of the periodic medium, the electromagnetic field $\mathbf{E}\left(\mathbf{r},\omega\right)$
surrounding the moving charge can be expressed in the form of the
Bloch eigenmode expansion (\ref{eq:E(r,w)_general}). This expansion
is valid for any point \textbf{$\mathbf{r}$} in the medium being
different from, but as close as required to, the charge trajectory.
Substituting the Fourier transform of the current density (\ref{eq:current_omega})
and the Bloch mode expansion (\ref{eq:E(r,w)_general}) in the equation
(\ref{eq:power_omega}) we obtain\begin{widetext}\begin{equation}
\frac{dP}{dl}=-\frac{1}{\left(2\pi\right)^{d}}\frac{\omega}{\pi\varepsilon_{0}}\sum_{n}\int_{BZ}d^{d}k\,\mathrm{Re}\left[-\mathrm{i}\left\{ \frac{I^{\left(T\right)}}{\left(\omega-\omega_{\mathbf{k}n}^{(T)}+\mathrm{i}\gamma\right)\left(\omega+\omega_{\mathbf{k}n}^{(T)}+\mathrm{i}\gamma\right)}+\frac{I^{\left(L\right)}}{\left(\omega+\mathrm{i}\gamma\right)^{2}}\right\} \right]\label{eq:power1}\end{equation}
with \begin{equation}
I^{\left(\alpha\right)}=q^{2}I_{1}^{\left(\alpha\right)}I_{2}^{\left(\alpha\right)}=q^{2}\left\{ \intop_{-\infty}^{\infty}dr_{\parallel}\,\left(\mathbf{e}_{\mathbf{k}n}^{(\alpha)\star}\left(r_{\parallel}\right)\cdot\hat{\mathbf{r}}_{\parallel}\right)\mathrm{e}^{-\mathrm{i}\left(k_{\parallel}-\frac{\omega}{\left|\mathbf{v}\right|}\right)r_{\parallel}}\right\} \left\{ \frac{1}{l}\intop_{-l/2}^{l/2}dr_{\parallel}\,\left(\mathbf{e}_{\mathbf{k}n}^{(\alpha)}\left(r_{\parallel}\right)\cdot\hat{\mathbf{r}}_{\parallel}\right)\mathrm{e}^{\mathrm{i}\left(k_{\parallel}-\frac{\omega}{\left|\mathbf{v}\right|}\right)r_{\parallel}}\right\} ,\label{eq:integrals}\end{equation}
\end{widetext}where $\alpha=T,\, L$. We have readily performed the
space integration in the transverse direction $\hat{\mathbf{r}}_{\perp}$
and used the Bloch theorem $\mathbf{E}_{\mathbf{k}n}^{(\alpha)}\left(\mathbf{r}\right)=\mathbf{e}_{\mathbf{k}n}^{(\alpha)}\left(\mathbf{r}\right)\exp\left(\mathrm{i}\mathbf{k}\cdot\mathbf{r}\right)$,
where $\mathbf{e}_{\mathbf{k}n}^{(\alpha)}\left(\mathbf{r}\right)$
is a lattice periodic function. To avoid having to deal with the {}``bremsstrahlung''
radiation we limit ourselves to the electron trajectories which do
not cut dielectric interfaces in the periodic medium. Such trajectories
are necessarily rationally oriented with respect to the periodic lattice.
In this case the function $\left(\mathbf{e}_{\mathbf{k}n}^{(\alpha)}\left(r_{\parallel}\right)\cdot\hat{\mathbf{r}}_{\parallel}\right)$
in (\ref{eq:integrals}) as well as its complex conjugate are both
one dimensional periodic functions with a period $a$ defined by a
particular orientation of the electron trajectory. Then, Eq. (\ref{eq:integrals})
can be further simplified to (see Appendix B) \begin{equation}
I^{\left(\alpha\right)}=2\pi q^{2}\sum_{m}\left|c_{m}^{\left(\alpha\right)}\left(\mathbf{k};n\right)\right|^{2}\delta\left(k_{\parallel}-\frac{\omega}{\left|\mathbf{v}\right|}-\frac{2\pi}{a}m\right).\label{eq:integrals_final}\end{equation}
Here $k_{\parallel}$ is the component of the wave vector parallel
to the electron trajectory. $c_{m}(\mathbf{k};n)$ is the $m$-th
($m\in\mathbb{Z}$) Fourier coefficient of the periodic function $\left(\mathbf{e}_{\mathbf{k}n}^{(\alpha)}\left(r_{\parallel}\right)\cdot\hat{\mathbf{r}}_{\parallel}\right)$
defined as \begin{equation}
c_{m}^{\left(\alpha\right)}\left(\mathbf{k};n\right)=\frac{1}{a}\int_{0}^{a}dr_{\parallel}\,\left(\mathbf{e}_{\mathbf{k}n}^{(\alpha)}\left(r_{\parallel}\right)\cdot\hat{\mathbf{r}}_{\parallel}\right)\mathrm{e}^{-\mathrm{i}\frac{2\pi}{a}mr_{\parallel}}.\label{eq:coefficient_Cm}\end{equation}

\begin{figure}[!b]
\begin{centering}
\includegraphics[width=0.9\columnwidth]{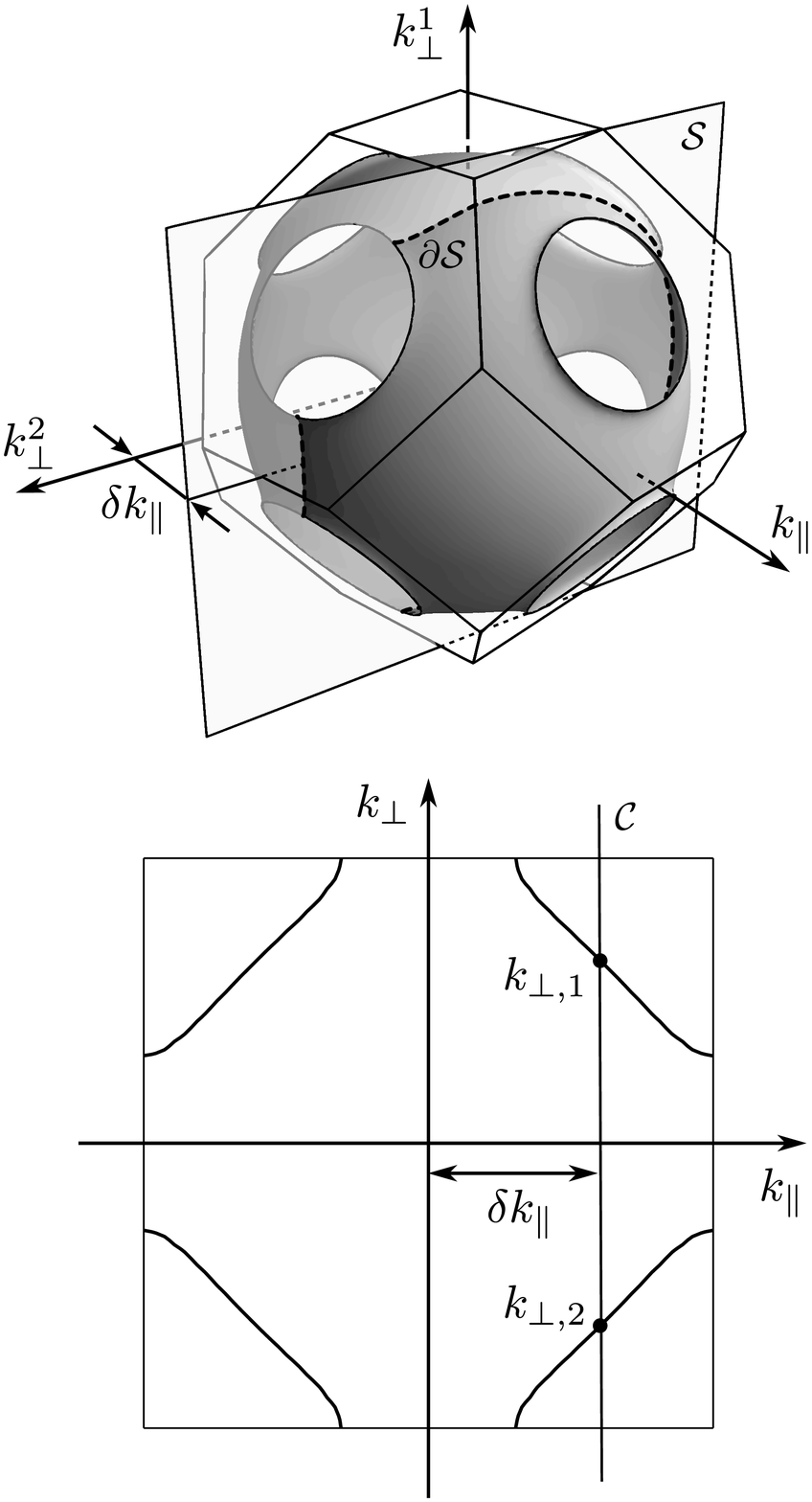}
\par\end{centering}

\caption{(Top) Diagram to define the integration plane $\mathcal{S}$ and the
integration contour $\partial\mathcal{S}$ (dashed line) in Eq.~(\ref{eq:power3},\ref{eq:power_final_3D}).
Iso-frequency surface enclosed in the first BZ of the FCC lattice
is shown for a normalized frequency $\omega_{\mathbf{k}n}=\omega$
inside the first bandgap of a 3D inverted opal \cite{rom03}. (Bottom)
Diagram to define the integration line $\mathcal{C}$ and the set
of points $\left\{ k_{\perp,i}\right\} $ (two thick dots) in Eqs.~(\ref{eq:power3},\ref{eq:power_final_2D}).
Iso-frequency contour enclosed in the first BZ of a square lattice
PhC is shown for a normalized frequency $\omega_{\mathbf{k}n}=\omega$
inside the first bandgap. The plane $\mathcal{S}$ and the line $\mathcal{C}$
are defined by the relation $k_{\parallel}=\delta k_{\parallel}=\frac{\omega}{\left|\mathbf{v}\right|}+\frac{2\pi}{a}m$.
The choice of the coordinate system with one axis, $k_{\parallel}$,
parallel to the electron trajectory is shown.\label{fig:Iso-surface}}

\end{figure}

Taking into account the expression (\ref{eq:integrals_final}) and
the relation $\mathrm{Re}\left[\mathrm{i}\left(\mathrm{Re\left[z\right]+\mathrm{i\,}Im\left[z\right]}\right)\right]=-\mathrm{Im}\left[z\right]$,
the power radiated by a moving charge per unit length is given by\begin{widetext}\begin{multline}
\frac{dP}{dl}=-\frac{1}{\left(2\pi\right)^{d-1}}\frac{\omega q^{2}}{\pi\varepsilon_{0}}\sum_{nm}\int_{BZ}d^{d}k\,\delta\left(k_{\parallel}-\frac{\omega}{\left|\mathbf{v}\right|}-\frac{2\pi}{a}m\right)\left\{ \left|c_{m}^{\left(T\right)}\left(\mathbf{k};n\right)\right|^{2}\mathrm{Im}\left[\frac{1}{\left(\omega-\omega_{\mathbf{k}n}^{(T)}+\mathrm{i}\gamma\right)\left(\omega+\omega_{\mathbf{k}n}^{(T)}+\mathrm{i}\gamma\right)}\right]+\right.\\
\left.\left|c_{m}^{\left(L\right)}\left(\mathbf{k};n\right)\right|^{2}\mathrm{Im}\left[\frac{1}{\left(\omega+\mathrm{i}\gamma\right)^{2}}\right]\right\} .\label{eq:power2}\end{multline}
This expression can be further integrated along the direction $k_{\parallel}$
in the $k$-space yielding \begin{multline}
\frac{dP}{dl}=-\frac{1}{\left(2\pi\right)^{d-1}}\frac{\omega q^{2}}{\pi\varepsilon_{0}}\sum_{nm}\int_{\mathcal{S}}d^{d-1}k_{\perp}\,\left\{ \left|c_{m}^{\left(T\right)}\left(\mathbf{k};n\right)\right|^{2}\mathrm{Im}\left[\frac{1}{\left(\omega-\omega_{\mathbf{k}n}^{(T)}+\mathrm{i}\gamma\right)\left(\omega+\omega_{\mathbf{k}n}^{(T)}+\mathrm{i}\gamma\right)}\right]+\right.\\
\left.\left|c_{m}^{\left(L\right)}\left(\mathbf{k};n\right)\right|^{2}\mathrm{Im}\left[\frac{1}{\left(\omega+\mathrm{i}\gamma\right)^{2}}\right]\right\} .\label{eq:power3}\end{multline}
\end{widetext}In the 3D case, a resulting surface integral is taken
over a plane $\mathcal{S}$. In the 2D case, the integration reduces
to an integral over a line $\mathcal{C}$ (Fig.~\ref{fig:Iso-surface}).
Both the integration plane and the integration line should be orthogonal
to the electron trajectory and are defined by the following relation
\begin{equation}
k_{\parallel}=\frac{\omega}{\left|\mathbf{v}\right|}+\frac{2\pi}{a}m.\label{eq:cut1}\end{equation}
Here the integer $m$ should be chosen in such a way that the wave
vector $k_{\parallel}$ stays in the first BZ. Further, taking the
limit $\gamma\rightarrow0^{+}$ and using the relation \begin{equation}
\mathrm{Im}\left[\underset{\gamma\rightarrow0^{+}}{\lim}\frac{1}{\omega\pm\omega_{\mathbf{k}n}^{(T)}+\mathrm{i}\gamma}\right]=-\pi\delta\left(\omega\pm\omega_{\mathbf{k}n}^{(T)}\right),\label{eq:cauchy}\end{equation}
the spectral radiated power (\ref{eq:power3}) can be expressed in
the form \begin{widetext}\begin{equation}
\frac{dP}{dl}=\frac{1}{\left(2\pi\right)^{d-1}}\frac{\omega q^{2}}{\pi\varepsilon_{0}}\sum_{nm}\int_{\mathcal{S}}d^{d-1}k_{\perp}\,\left\{ \frac{\pi}{2\omega_{\mathbf{k}n}^{(T)}}\left|c_{m}^{\left(T\right)}\left(\mathbf{k};n\right)\right|^{2}\left(\delta\left(\omega-\omega_{\mathbf{k}n}^{(T)}\right)-\delta\left(\omega+\omega_{\mathbf{k}n}^{(T)}\right)\right)+\frac{2\pi}{\omega}\left|c_{m}^{\left(L\right)}\left(\mathbf{k};n\right)\right|^{2}\delta\left(\omega\right)\right\} .\label{eq:power4}\end{equation}

The eigenfrequencies of the Bloch modes are positive \cite{jmw95},
so the second term in Eq. (\ref{eq:power4}) containing the delta
function $\delta\left(\omega+\omega_{\mathbf{k}n}^{(T)}\right)$ is
zero for all frequencies. The third term in Eq. (\ref{eq:power4})
is due to the work the current does on the longitudinal part of the
electromagnetic field. In the presence of free charges the longitudinal
part of the field corresponds to the static electric field and the
work done against it results in non-radiative energy transfer with
a nonzero contribution only at zero frequency. In what follows we
will disregard this non-radiative contribution and will limit ourselves
to the radiation into propagating electromagnetic waves only. Then
the spectral radiated power is given by\begin{equation}
\frac{dP}{dl}=\frac{1}{\left(2\pi\right)^{d-2}}\frac{\omega q^{2}}{4\pi\varepsilon_{0}}\sum_{nm}\int_{\mathcal{S}}d^{d-1}k_{\perp}\,\frac{\left|c_{m}^{\left(T\right)}\left(\mathbf{k};n\right)\right|^{2}\delta\left(\omega-\omega_{\mathbf{k}n}^{(T)}\right)}{\omega_{\mathbf{k}n}^{(T)}}\label{eq:power5}\end{equation}
\end{widetext}

\begin{figure}[b]
\begin{centering}
\includegraphics[width=1\columnwidth]{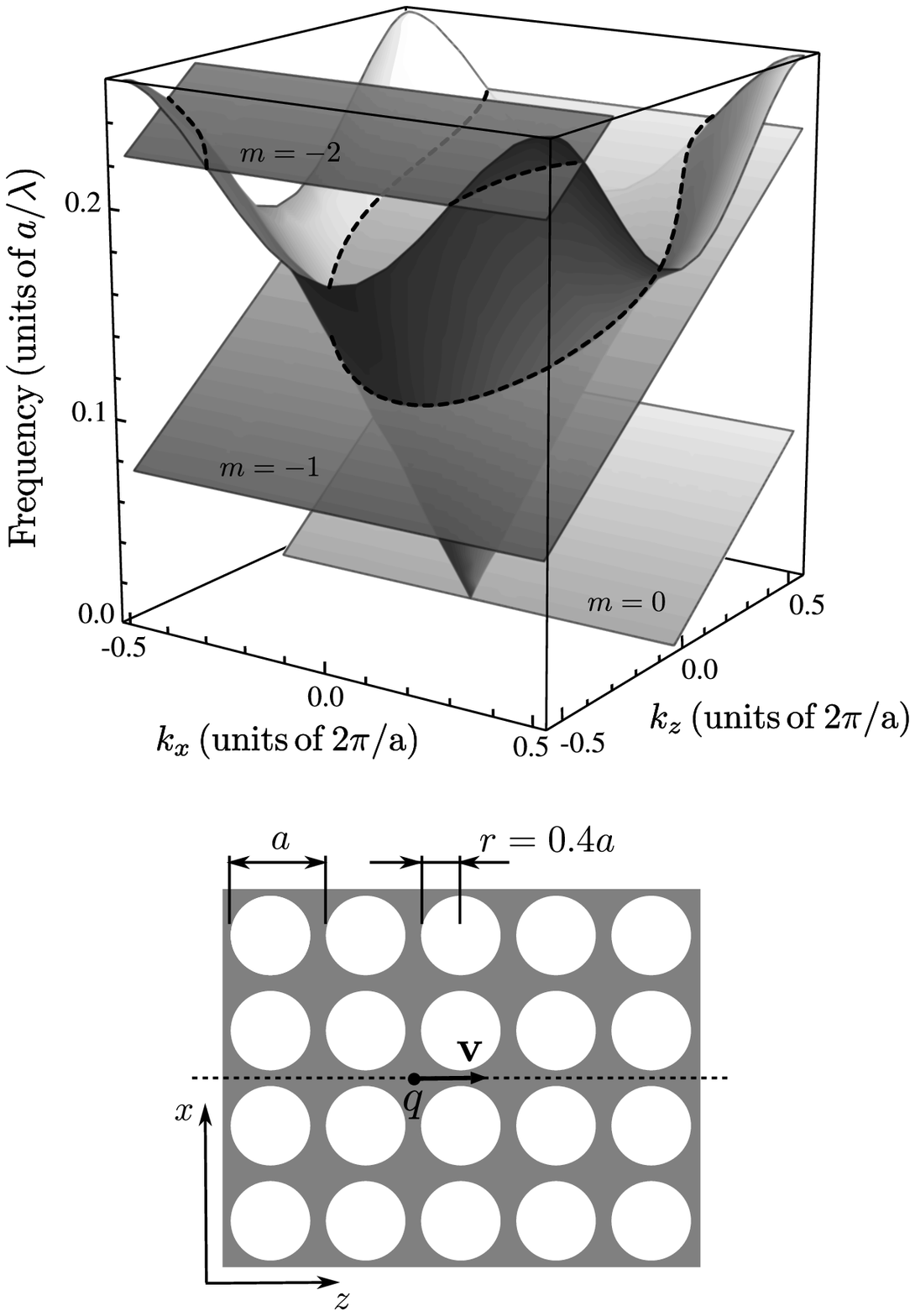}
\par\end{centering}

\caption{Diagram to illustrate the generalized Cherenkov condition (\ref{eq:cherenkov_condition_periodic}).
A 3D representation of the photonic band structure (top) of the 2D
PhC (bottom) is shown for TE polarization. An infinite 2D square lattice
of air holes in a dielectric medium is considered. The radius of the
holes is $r=0.4a$, the dielectric constant of the background medium
is $\varepsilon=12.0$. Only the first band in the first BZ is presented.
The right-hand side of Eq. (\ref{eq:cherenkov_condition_periodic})
defines the set of planes for different $m$. The intersection (dashed line)
of these planes with the band structure  determines the Bloch
modes contributing to the Cherenkov radiation. Here it is supposed
that a line charge moves along the $z$-axis in the crystal with
velocity $\left|\mathbf{v}\right|=0.15c$.\label{fig:cherenkov-condition}}

\end{figure}

The argument of the Dirac delta function in Eq. (\ref{eq:power5})
is a function of the wave vector. One can use this fact to further reduce
the dimensionality of the $\left(d-1\right)$ $k$-space integral.
In the 3D case, using the relation \begin{equation}
\intop_{\mathcal{V}}d^{d}k\, f\left(\mathbf{k}\right)\delta\left(g\left(\mathbf{k}\right)\right)=\intop_{\partial\mathcal{V}}d^{d-1}k\,\frac{f\left(\mathbf{k}\right)}{\left|\nabla_{\mathbf{k}}g\left(\mathbf{k}\right)\right|},\label{eq:grad_dD}\end{equation}
where $\partial\mathcal{V}$ is $\left(d-1\right)$ dimensional surface
defined by $g\left(\mathbf{k}\right)=0$, the integral over the plane
$\mathcal{S}$ is converted into a contour integral\begin{equation}
\left(\frac{dP}{dl}\right)^{3D}=\frac{q^{2}}{8\pi^{2}\varepsilon_{0}}\sum_{nm}\int_{\partial\mathcal{S}}dk\,\frac{\left|c_{m}^{\left(T\right)}(\mathbf{k};n)\right|^{2}}{\left|\nabla_{\mathbf{k}_{\bot}}\omega_{\mathbf{k}n}^{(T)}\right|}.\label{eq:power_final_3D}\end{equation}
The contour $\mathcal{\partial\mathcal{S}}$ is defined by the relation
(\ref{eq:cut1}) and \begin{equation}
\omega_{\mathbf{k}n}^{(T)}=\omega.\label{eq:cut2}\end{equation}
It is an intersection of the iso-frequency surface with plane $\mathcal{S}$
(Fig. \ref{fig:Iso-surface}-top). In the 2D case, the relation \begin{equation}
\delta\left(f\left(k\right)\right)=\sum_{i}\frac{\delta\left(k-k_{i}\right)}{\left|f'\left(k_{i}\right)\right|}\label{eq:grad_1D}\end{equation}
can be used, where summation is taken over all solutions of the equation
$f\left(k\right)=0$. Substituting (\ref{eq:grad_1D}) into (\ref{eq:power5})
we obtain\begin{equation}
\left(\frac{dP}{dl}\right)^{2D}=\frac{\omega q^{2}}{4\pi\varepsilon_{0}}\sum_{nmi}\int_{\mathcal{C}}dk_{\perp}\,\frac{\left|c_{m}^{\left(T\right)}\left(\mathbf{k};n\right)\right|^{2}\delta\left(k_{\perp}-k_{\perp,i}\right)}{\omega_{\mathbf{k}n}^{(T)}\left.\left(\left|\partial\omega_{\mathbf{k}n}^{(T)}/\partial k_{\perp}\right|\right)\right|_{k_{\perp,i}}},\label{eq:power_2D}\end{equation}
where $\left\{ k_{\perp,i}\right\} $ are simultaneous solutions of
the equations (\ref{eq:cut1}) and (\ref{eq:cut2}) given by the intersections
of the iso-frequency contour with the line $\mathcal{C}$ (Fig. \ref{fig:Iso-surface}-bottom).
Performing $k$-space integration, we finally obtain\begin{equation}
\left(\frac{dP}{dl}\right)^{2D}=\frac{q^{2}}{4\pi\varepsilon_{0}}\sum_{nmi}\left.\left(\frac{\left|c_{m}^{\left(T\right)}\left(\mathbf{k};n\right)\right|^{2}}{\left|\partial\omega_{\mathbf{k}n}^{(T)}/\partial k_{\perp}\right|}\right)\right|_{k_{\perp,i}},\label{eq:power_final_2D}\end{equation}
where the function in brackets is calculated for the wave vectors
corresponding to the set $\left\{ k_{\perp,i}\right\} $.

Formulas (\ref{eq:power_final_3D}) and (\ref{eq:power_final_2D})
constitute the main result of the present work. They give the power
radiated by the moving point charge (3D) or line charge (2D) $q$ 
in the spectral interval
$\left[\omega,\,\omega+d\omega\right]$ per unit length of the trajectory
for a 3D and 2D periodic medium, respectively. The radiated power
is proportional to the Fourier coefficients $c_{m}^{\left(T\right)}(\mathbf{k};n)$,
which effectively describe the local coupling strength between the
current density produced by a moving charge and the electromagnetic
field at the electron location. The gradient and derivative of the
dispersion relation $\mathbf{v}_{\perp}^{g}=\nabla_{\mathbf{k}_{\bot}}\omega_{\mathbf{k}n}^{(T)}$
and $v_{\perp}^{g}=\partial\omega_{\mathbf{k}n}^{(T)}/\partial k_{\perp}$
yield the component of the group velocity, $\mathbf{v}^{g}$, of the
Bloch eigenmode $\left(\mathbf{k};n\right)$, which is orthogonal
to the electron trajectory. The Cherenkov radiated power is proportional
to the inverse of this component of the group velocity. That means
that the radiated power can be \emph{strongly enhanced} not only if
the group velocity itself is small for some frequency, but also if
the component of the group velocity orthogonal to the electron trajectory
becomes small. At the same time \emph{suppression} of the Cherenkov
radiation is possible if for some frequency the current density produced
by a moving charge is not coupled to the corresponding Bloch mode
and the Fourier coefficients $c_{m}^{\left(T\right)}(\mathbf{k};n)$
is small.

Only eigenmodes with the wave vectors on the contour $\partial\mathcal{S}$
(\ref{eq:power_final_3D}) and from the set $\left\{ k_{\perp,i}\right\} $
(\ref{eq:power_final_2D}) contribute to the radiated power at a given
frequency. It is important to realize that Eqs. (\ref{eq:cut1}) and
(\ref{eq:cut2}) defining the contour $\partial\mathcal{S}$ and the
set $\left\{ k_{\perp,i}\right\} $ are equivalent to the Cherenkov
resonance condition (\ref{eq:cherenkov-condition}). In fact, substituting
(\ref{eq:cut2}) in (\ref{eq:cut1}) and taking into account that
the scalar product in (\ref{eq:cherenkov-condition}) results in $\mathbf{v}\cdot\mathbf{k}=\left|\mathbf{v}\right|k_{\parallel}$
one obtains \emph{the generalized Cherenkov condition} for a periodic
medium\begin{equation}
\omega_{\mathbf{k}n}^{(T)}=\left|\mathbf{v}\right|k_{\parallel}-\left|\mathbf{v}\right|\frac{2\pi}{a}m.\label{eq:cherenkov_condition_periodic}\end{equation}
In the 4D (3D) ($\omega$-$k$)-space the right-hand side of the relation
(\ref{eq:cherenkov_condition_periodic}) defines a hyperplane (plane)
whose intersection with the band structure, $\omega_{\mathbf{k}n}^{(T)}$,
determines Bloch modes contributing to the Cherenkov radiation (Fig.
\ref{fig:cherenkov-condition}-top). Nonzero integers $m$ ensure
that such an intersection and consequently the Cherenkov radiation
exist in a periodic medium for an \emph{arbitrarily small} charge
velocity. In a homogeneous medium $m=0$ and the Cherenkov condition
reduces to a standard form $\omega_{\mathbf{k}}=\left|\mathbf{v}\right|k_{\parallel}$. 

As a simple check of our theory we show in the following that the
final formulas (\ref{eq:power_final_3D},\ref{eq:power_final_2D})
reproduce the limit of a homogeneous medium with the dielectric constant
$\varepsilon$. For a given frequency $\omega$, the wave vector $\mathbf{\left|k\right|}$
and the group velocity $\left|\mathbf{v}_{\perp}^{g}\right|$ are
given by $\mathbf{\left|k\right|}=\left(\omega\sqrt{\varepsilon}\right)/c$
and $\left|\mathbf{v}_{\perp}^{g}\right|=\left(c\left|\mathbf{k}_{\perp}\right|\right)/\left(\sqrt{\varepsilon}\left|\mathbf{k}\right|\right)$
respectively, with $\mathbf{k}_{\perp}=\mathbf{k}-\mathbf{k}_{\parallel}$
being the component of the wave vector perpendicular to the electron
trajectory. The appropriately normalized eigenmodes are plane waves
$\mathbf{E}=\left(1/\sqrt{\varepsilon}\right)\mathbf{\hat{e}}\exp\left(i\mathbf{k}\cdot\mathbf{r}\right)$,
where $\mathbf{\hat{e}}$ is a polarization unit vector orthogonal
to the wave vector $\mathbf{k}$. Further, according to the Eq. (\ref{eq:cut1})
the wave vector component $k_{\parallel}$ is equal to $k_{\parallel}=\omega/\left|\mathbf{v}\right|$
with $m=0$ and the coefficient $c_{0}$ is given by $c_{0}=\left|\mathbf{k}_{\perp}\right|/\left(\sqrt{\varepsilon}\left|\mathbf{k}\right|\right)$.
Then in the 3D case, taking into account that an integration contour
$\partial\mathcal{S}$ is a circle of radius $\left|\mathbf{k}_{\perp}\right|$
and performing integration in polar coordinates with $dk=\left|\mathbf{k}_{\perp}\right|d\phi$,
the radiated power (\ref{eq:power_final_3D}) is given by \begin{equation}
\left(\frac{dP}{dl}\right)_{h}^{3D}=\frac{1}{4\pi\varepsilon_{0}}\frac{q^{2}}{c\sqrt{\varepsilon}}\left|\mathbf{k}\right|\left(1-\left(\frac{\left|\mathbf{k}_{\parallel}\right|}{\left|\mathbf{k}\right|}\right)^{2}\right),\label{eq:frank_tamm}\end{equation}
which finally yields the usual results of the Frank-Tamm theory \cite{ft37}\begin{equation}
\left(\frac{dP}{dl}\right)_{h}^{3D}=\frac{q^{2}\omega}{4\pi\varepsilon_{0}c^{2}}\left(1-\frac{c^{2}}{\varepsilon\left|\mathbf{v}\right|^{2}}\right).\label{eq:frank_tamm_3D}\end{equation}
In the 2D case Eq. (\ref{eq:power_final_2D}) yields\begin{equation}
\left(\frac{dP}{dl}\right)_{h}^{2D}=\frac{1}{2\pi\varepsilon_{0}}\frac{q^{2}}{c\sqrt{\varepsilon}}\sqrt{1-\frac{c^{2}}{\varepsilon\left|\mathbf{v}\right|^{2}}}.\label{eq:frank_tamm_2D}\end{equation}

\section{Numerical results\label{sec:Numerical-example}}

In this section the analytical results developed in the previous section
are applied to the numerical study of the Cherenkov radiation in a
2D photonic crystal. An infinite 2D square lattice of air holes in
a dielectric medium is considered. The radius of the holes is $r=0.4a$,
while the dielectric constant of the background medium is $\varepsilon=12.0$.
A line charge oriented perpendicular to the periodicity plane of the
crystal moves along the $z$-axis with a velocity $v$, staying always
in the space between air holes (Fig. \ref{fig:cherenkov-condition}
bottom). The corresponding current density, Eqs. (\ref{eq:current_time})
and (\ref{eq:current_omega}), generates an electric field (\ref{eq:E(r,w)_general})
polarized in the periodicity plane (transverse electric or TE polarization)
\cite{jmw95,sak01}, i.e., the Bloch eigenmode expansion should include
the TE polarized Bloch modes only. The first TE band for the considered
PhC is presented in the figure~\ref{fig:cherenkov-condition}-top.
The band structure was calculated using the plane wave expansion method
\cite{jj01}.

\begin{figure}
\begin{centering}
\includegraphics[width=1\columnwidth]{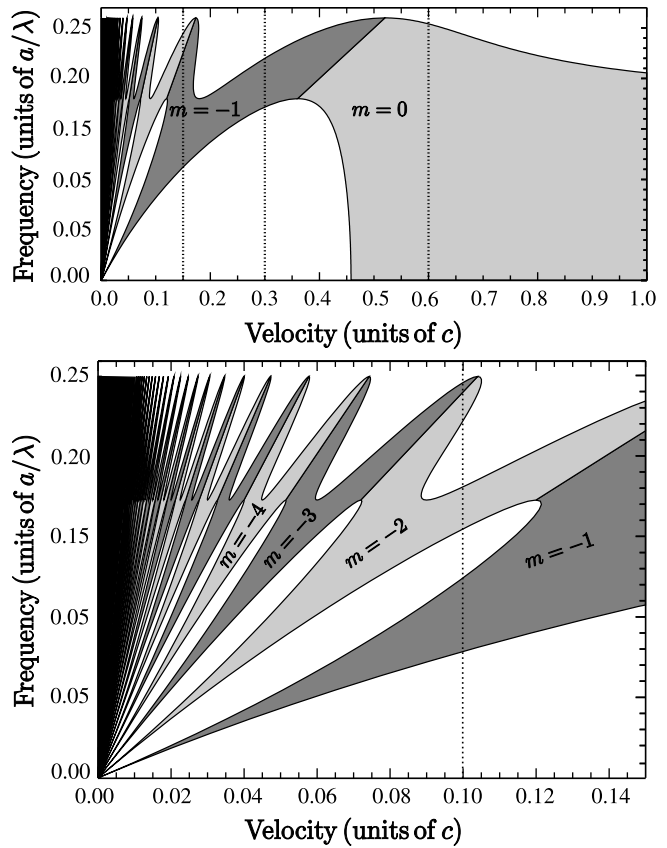}
\par\end{centering}

\caption{Cherenkov radiation band. Sub-bands defined by the intersections of
the band structure with the planes corresponding to the different
$m$'s are shaded in light and dark gray. Vertical lines mark the
charge velocities used in the further calculations, $v=0.15c$, $v=0.3c$
and $v=0.6c$ in the top panel and $v=0.1c$ in the bottom panels,
respectively.\label{fig:cherenkov_bands}}

\end{figure}
\begin{figure}
\begin{centering}
\includegraphics[width=1\columnwidth]{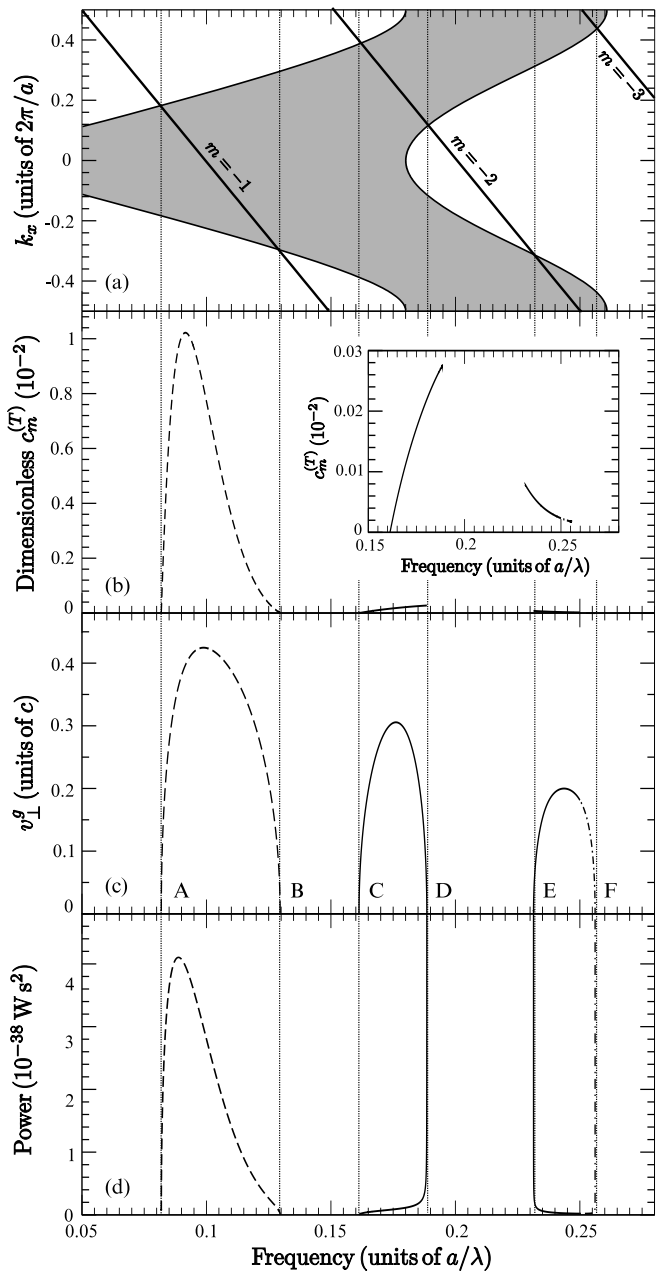}
\par\end{centering}

\caption{Cherenkov emission spectrum for the charge velocity $v=0.1c$. Projections
of the first band of the considered photonic crystal and the planes
$m=-1$, $m=-2$ and $m=-3$ defining the Cherenkov band on the $k_{x}$-$\omega$
plane are shown (a). The Fourier coefficients, $c_{m}^{\left(T\right)}\left(\mathbf{k};n\right)$,
the orthogonal component of the group velocity, $v_{\perp}^{g}$,
and the Cherenkov power spectrum are shown in panels (b), (c) and
(d), respectively. Vertical lines mark the edges of the Cherenkov
sub-bands. Contribution from the sub-bands corresponding to $m=-1$,
$m=-2$ and $m=-3$ are shown as dashed, solid and dashed-dotted lines,
respectively. \label{fig:spectrum_01}}

\end{figure}

To find the power radiated by a charge moving with a given velocity
$v$, all Bloch modes contributing to the radiation should be determined.
These modes are specified by the solutions of the relation (\ref{eq:cherenkov_condition_periodic}).
In what follows we restrict our analysis to the frequency range of
the first band of the considered PhC structure. In figure \ref{fig:cherenkov-condition}-top,
solutions of the Cherenkov relations (\ref{eq:cherenkov_condition_periodic})
are graphically illustrated for $m=0,\,-1,\,-2$ (dashed lines) and
charge velocity $v=0.15c$. The frequencies satisfying relations (\ref{eq:cherenkov_condition_periodic})
determine the spectral range of nonzero contribution to the Cerenkov
radiated power, \emph{the Cherenkov band}. The evolution of the Cherenkov
band is presented in figure~\ref{fig:cherenkov_bands} as a function
of the charge velocity.

For charge velocity $v=0.15c$, the Cherenkov spectrum is given by
the intersections of the band structure with the planes $m=-1$ and
$m=-2$. The plane corresponding to $m=0$ does not intersect the
band structure of the crystal (Figs.~\ref{fig:cherenkov-condition},~\ref{fig:cherenkov_bands}).
For smaller charge velocities, more and more planes intersect the
photonic band structure, and the Cherenkov spectrum is built from a number of
discrete sub-bands. For sufficiently small charge velocities the spectral
range of the first photonic band becomes densely filled with the discrete
sub-bands (Fig.~\ref{fig:cherenkov_bands}).

In the long wavelength limit the periodic medium is effectively homogeneous.
For the PhC considered in Fig.~\ref{fig:cherenkov_bands} the effective refractive index is equal to $n_{\mathrm{eff}}=\sqrt{\varepsilon_{\mathrm{eff}}}\thickapprox2.186$.
Consequently, for $m=0$ the relation (\ref{eq:cherenkov_condition_periodic})
imposes a condition on the minimal charge velocity to produce 
Cherenkov radiation at small frequencies, namely $v\geq v_{\mathrm{min}}=
c/n_{\mathrm{eff}}\approx0.457c$.
For charge velocities larger than the threshold value the Cherenkov
band covers the spectral range from zero to the maximum frequency,
which is defined by the intersection of the band structure with the
plane $m=0$ at the first BZ boundary (Fig.~\ref{fig:cherenkov_bands}).

To compute the radiated power from Eq.~(\ref{eq:power_final_2D}), 
one should calculate
the Bloch modes along a charge trajectory, their Fourier transforms
and corresponding group velocities for wave vectors belonging to the
intersections defined by Eq. (\ref{eq:cherenkov_condition_periodic}).
The calculation of the Cherenkov spectrum is illustrated in figure~\ref{fig:spectrum_01}
for a line charge ($q=1.6\times10^{-19}\, C$) moving with the velocity
$v=0.1c$. To calculate Bloch modes and group velocities, the plane
wave expansion method \cite{jj01} and the Hellmann-Feynman theorem
were used, respectively. For the velocity $v=0.1c$ the Cherenkov
spectrum consists of three sub-bands defined by the planes $m=-1$, $m=-2$
and $m=-3$ (Figs.\ref{fig:cherenkov_bands}, \ref{fig:spectrum_01}),
respectively. The Fourier coefficients, $c_{m}^{\left(T\right)}\left(\mathbf{k};n\right)$,
(Fig.~\ref{fig:spectrum_01}-b) and the orthogonal component of the
group velocity, $v_{\perp}^{g}$, (Fig.~\ref{fig:spectrum_01}-c)
are nonzero only within the sub-bands. Both Fourier coefficients and
the orthogonal component of the group velocity approach zero at sub-band
edges A, B and C. At the edges D, E, and F only the orthogonal component
of the group velocity is zero, while the Fourier coefficients have
finite nonzero value. Calculation of the Cherenkov radiated power
at the band edges A, B and C leads to the indeterminate limits of
the form $0/0$, which can be evaluated using l'Hopital's rule and
is equal to zero. At the band edges D, E and F the Cherenkov power diverges in
an intgegrable way.

\begin{figure}
\begin{centering}
\includegraphics[width=0.85\columnwidth]{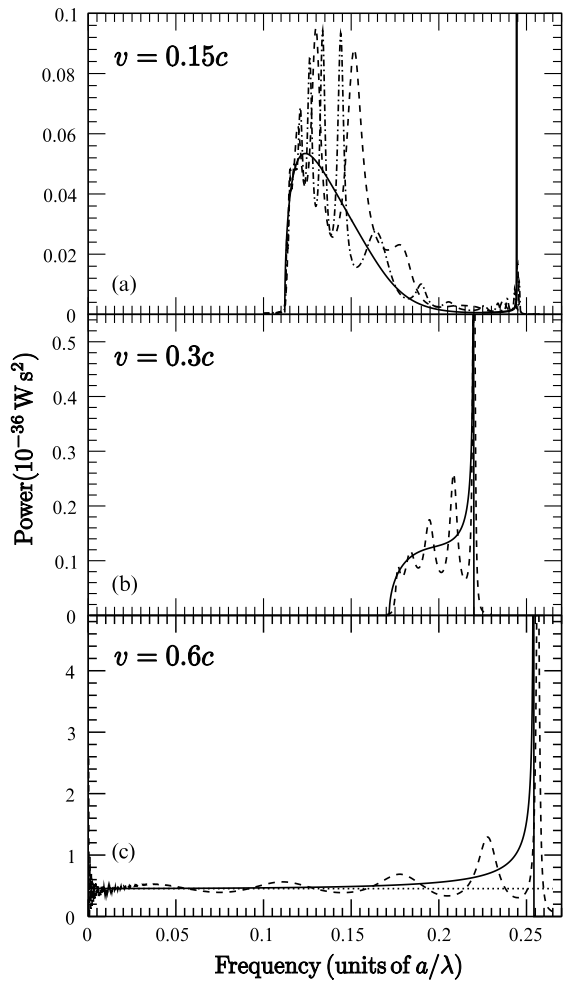}
\par\end{centering}

\caption{Cherenkov emission spectra for different charge velocities in a 2D
photonic crystal. Solid lines correspond to the analytical results
(\ref{eq:power_final_2D}). Dotted line in the bottom panel ($v=0.6c$)
corresponds to the Cherenkov radiated power in the homogeneous medium
with $n_{\mathrm{eff}}=2.186$, Eq. (\ref{eq:frank_tamm_2D}). Radiated
power spectra obtained using FDTD method are shown as dashed lines.
For charge velocity $v=0.15c$ radiated power spectra are show for
both $10a\times188a$ (dashed line) and $20a\times188a$ (dashed-dotted
line) structures. \label{fig:Emission-spectra}}

\end{figure}

In figure~\ref{fig:Emission-spectra} the Cherenkov radiated power
spectra are shown for charge velocities $v=0.15c$, $v=0.3c$ and
$v=0.6c$. For charge velocities smaller than the threshold value
$v_{\mathrm{min}}\approx0.457c$ the Cherenkov radiation is nonzero
only within single or multiple spectral bands. For velocities above
the threshold, the radiated power is nonzero almost everywhere within
the first band, approaching asymptotically the value of the Cherenkov
radiated power in a homogeneous medium with $n=n_{\mathrm{eff}}$
for small frequencies. The radiated power calculated using Eq. (\ref{eq:frank_tamm_2D})
for $v=0.6c$ and $n_{\mathrm{eff}}=2.186$ is shown in figure~\ref{fig:Emission-spectra}
(bottom panel) as dotted line. The Cherenkov radiated power is enhanced
near those frequencies where the group velocity component orthogonal
to the charge trajectory vanishes, while the Fourier coefficients
remain finite (Fig.~\ref{fig:Emission-spectra}).

To substantiate our analytical results the direct numerical integration
of the Maxwell's equations has been performed using rigorous finite-difference
time-domain (FDTD) method \cite{taf95}. The simulated structure
was a $10a\times Na$ lattice of air holes in a homogeneous medium
with $\varepsilon=12.0$. The longitudinal size of the periodic
structure was set to $N=188$, $N=376$ and $N=752$ lattice constants
for an charge velocity $v=0.15c$, $v=0.3c$ and $v=0.6c$, respectively.
The lattice was surrounded by a $2a$ wide layer of homogeneous material.
The simulation domain was discretized into squares with a side $\Delta=a/18$
and was surrounded by a $35$-cell-wide perfectly matched layer (PML)
\cite{ber94}. The time step of integration was set to  $98\%$ of
the Courant value. The moving line source (\ref{eq:current_time})
was modeled as a current density source \cite{taf95} with the Dirac
delta function represented via an appropriately normalized Kronecker
delta $\delta_{ij}/\Delta^{2}$. The charge trajectory was oriented 
in the longitudinal direction of the system, 
placed in the geometrical center of
the crystal, exactly between the 5th and the 6th row of holes.

For the numerical computation of the radiated power the electric and 
magnetic fields were
stored at a detector surface enclosing the crystal,
 and their Fourier transforms with respect to 
time were found by
discrete Fourier transformation. The longitudinal dimension of the
structure was different for different charge velocities in order to
keep the integration time at the detector and consequently the spectral
resolution constant. The detector surface was situated
in the close vicinity of the crystal boundary. The total radiated
power per unit length was then calculated as
\begin{equation}
\frac{dP}{dl}=\frac{1}{d}\frac{2}{\pi}\intop_{0}^{D}dz\,\mathbf{S}\left(z,\omega\right)\cdot\mathbf{\hat{n}}\label{eq:power-fdtd}
\end{equation}
where $\mathbf{S}\left(z,\omega\right)=\frac{1}{2}\mathrm{Re}\left[\mathbf{E}\left(z,\omega\right)\times\mathbf{H}^{\star}\left(z,\omega\right)\right]$
is the Poynting vector at radiation frequency $\omega$, $D$ is the length of
the detector surface
along the trajectory and $\hat{\mathbf{n}}$ is a unit vector orthogonal
to the detector interface.

An overall very good agreement between the results of the analytical
(Fig.~\ref{fig:Emission-spectra}, solid lines) and numerical calculations
(Fig.~\ref{fig:Emission-spectra}, dashed lines) is obtained. Both
the spectral range of a nonzero radiated power and its absolute value
are well represented using the FDTD method. The main difference can
be traced back to Fabry-Perot-like oscillations of the radiated power
due to the finite-size effects in the FDTD calculations. In the finite
structure, the Cherenkov radiated power stays considerably enhanced
near the band edges having large but finite value. The total power
oscillates around the analytical value becoming partially suppressed
or enhanced for different frequencies. To confirm that these oscillations
indeed result from the finite transverse dimension of the considered
photonic crystal, we have performed simulations for the crystal with
a double thickness ($20a\times188a$) for the charge velocity $v=0.15c$.
Resulting radiated power spectrum is shown in the corresponding panel
in figure~\ref{fig:Emission-spectra} as a dashed-dotted curve. One
can see twice as many oscillations as in the case of the thinner structure,
which is a typical signature of the Fabry-Perot like phenomena. The
further enhancement of the radiated power in comparison to the infinite
structure can be associated with the longer interaction time of the
charge at resonance frequencies with the effectively slow Fabry-Perot
modes of a photonic crystal slab.

\section{Conclusions\label{sec:Conclusions}}

In this paper, analytical expression for the Cherenkov power emitted
per unit length of the charge trajectory in the case of a general
3D and 2D periodic dielectric medium has been derived. The obtained
formula for the Cherenkov power involves the calculations of Bloch
modes and corresponding group velocities at a limited number of points of the
reciprocal space only, making the application of the proposed method
computationally not demanding. All calculations have been performed
on a desktop PC and our method requires 5 to 10 times less CPU time
then FDTD method. The analysis of the Cherenkov emission spectrum
in the periodic medium reveals that the Cherenkov effect indeed exists
for every electron velocity. Similar to the case of the modification
of the dipole emission in a photonic crystal, the Cherenkov radiation
can be suppressed if the coupling of the current density produced
by a moving electron with a Bloch mode is poor. At the same time,
an enhancement of the Cherenkov radiation is possible also if only
the component of the group velocity orthogonal to the electron trajectory
is small. We have illustrated the developed analytical method and
its conclusions using a numerically rigorous finite-difference time-domain
method in a special case of a 2D photonic crystal and demonstrated
a reasonable agreement between numerical and analytical results.
\begin{acknowledgments}
We are grateful to Sergei Zhukovsky for fruitful discussions and careful
reading of the manuscript. Financial support from the Deutsche Forschungsgemeinschaft
(DFG FOR 557) is gratefully acknowledged.
\end{acknowledgments}

\section*{Appendix A}

Using the Fourier representation of the time depended real function
$\mathbf{F}\left(\mathbf{r},t\right)$, \begin{eqnarray}
\mathbf{F}\left(\mathbf{r},t\right) & = & \mathrm{Re}\left[\frac{1}{2\pi}\int_{-\infty}^{\infty}d\omega\,\mathbf{F}\left(\mathbf{r},\omega\right)\mathrm{e}^{-\mathrm{i}\omega t}\right]\label{eq:a1}\\
 & = & \frac{1}{4\pi}\int_{-\infty}^{\infty}d\omega\,\left(\mathbf{F}\left(\mathbf{r},\omega\right)\mathrm{e}^{-\mathrm{i}\omega t}+\mathbf{F}^{\star}\left(\mathbf{r},\omega\right)\mathrm{e}^{\mathrm{i}\omega t}\right),\nonumber \end{eqnarray}
for the electric field $\mathbf{E}\left(\mathbf{r},t\right)$ and
the current density $\mathbf{J}\left(\mathbf{r},t\right)$ in Eq.
(\ref{eq:power_time}), the total radiated energy (\ref{eq:total_energy})
can be written in the form\begin{eqnarray}
U & = & -\frac{2}{\left(4\pi\right)^{2}}\mathrm{Re}\left[\intop_{V_{0}}d^{d}r\,\int_{-\infty}^{\infty}dt\,\int_{-\infty}^{\infty}d\omega\,\int_{-\infty}^{\infty}d\Omega\,\right.\times\nonumber \\
 &  & \left\{ \mathbf{J}\left(\mathbf{r},\omega\right)\cdot\mathbf{E}\left(\mathbf{r},\Omega\right)\mathrm{e}^{-\mathrm{i}\left(\omega+\Omega\right)t}\right.+\label{eq:a2}\\
 &  & \left.\left.\mathbf{J}\left(\mathbf{r},\omega\right)\cdot\mathbf{E}^{\star}\left(\mathbf{r},\Omega\right)\mathrm{e}^{-\mathrm{i}\left(\omega-\Omega\right)t}\right\} \right].\nonumber \end{eqnarray}
Changing the integration order and using the integral relation for
the Dirac delta function\begin{equation}
\delta\left(x\right)=\frac{1}{2\pi}\int_{-\infty}^{\infty}dy\,\mathrm{e}^{-\mathrm{i}xy}\label{eq:a3}\end{equation}
 we obtain for total radiated energy\begin{eqnarray}
U & = & -\frac{1}{4\pi}\mathrm{Re}\left[\intop_{V_{0}}d^{d}r\,\int_{-\infty}^{\infty}d\omega\,\int_{-\infty}^{\infty}d\Omega\,\right.\times\nonumber \\
 &  & \left\{ \mathbf{J}\left(\mathbf{r},\omega\right)\cdot\mathbf{E}\left(\mathbf{r},\Omega\right)\delta\left(\omega+\Omega\right)\right.+\label{eq:a4}\\
 &  & \left.\left.\mathbf{J}\left(\mathbf{r},\omega\right)\cdot\mathbf{E}^{\star}\left(\mathbf{r},\Omega\right)\delta\left(\omega-\Omega\right)\right\} \right].\nonumber \end{eqnarray}
Further, integrating over $\Omega$ and using the symmetry of the
Fourier transform of the electric field, $\mathbf{E}\left(\mathbf{r},-\omega\right)=\mathbf{E}^{\star}\left(\mathbf{r},\omega\right)$,
the total radiated energy can be written as an integral over frequency
(\ref{eq:total_energy_omega})\begin{equation}
U=-\frac{2}{\pi}\int_{0}^{\infty}d\omega\,\frac{1}{2}\mathrm{Re}\left[\intop_{V_{0}}d^{d}r\,\mathbf{J}\left(\mathbf{r},\omega\right)\cdot\mathbf{E^{\star}}\left(\mathbf{r},\omega\right)\right].\label{eq:a5}\end{equation}
The integrand in Eq. (\ref{eq:a5}) coincides with the time-averaged
radiated power of the monochromatic source $\mathbf{J}\left(\mathbf{r},\omega\right)$
\cite{jac75}.

\section*{Appendix B}

Expanding a periodic function $\left(\mathbf{e}_{\mathbf{k}n}^{(\alpha)}\left(r_{\parallel}\right)\cdot\hat{\mathbf{r}}_{\parallel}\right)$
in the Fourier series\begin{equation}
\left(\mathbf{e}_{\mathbf{k}n}^{(\alpha)}\left(r_{\parallel}\right)\cdot\hat{\mathbf{r}}_{\parallel}\right)=\sum_{m=-\infty}^{\infty}c_{m}^{(\alpha)}\left(\mathbf{k};n\right)\mathrm{e}^{\mathrm{i}\frac{2\pi}{a}mr_{\parallel}}\label{eq:b1}\end{equation}
with coefficients $c_{m}^{\left(\alpha\right)}\left(\mathbf{k};n\right)$
defined in Eq. (\ref{eq:coefficient_Cm}), integrals $I_{1}^{\left(\alpha\right)}$
and $I_{2}^{\left(\alpha\right)}$ in (\ref{eq:integrals}) can be
transformed to \begin{equation}
I_{1}^{\left(\alpha\right)}=\sum_{m=-\infty}^{\infty}c_{m}^{(\alpha)\star}\left(\mathbf{k};n\right)\intop_{-\infty}^{\infty}dr_{\parallel}\,\mathrm{e}^{-\mathrm{i}\left(k_{\parallel}-\frac{\omega}{\left|\mathbf{v}\right|}-\frac{2\pi}{a}m\right)r_{\parallel}}\label{eq:b2}\end{equation}
and \begin{equation}
I_{2}^{\left(\alpha\right)}=\sum_{p=-\infty}^{\infty}c_{p}^{(\alpha)}\left(\mathbf{k};n\right)\frac{1}{l}\intop_{-l/2}^{l/2}dr_{\parallel}\,\mathrm{e}^{\mathrm{i}\left(k_{\parallel}-\frac{\omega}{\left|\mathbf{v}\right|}-\frac{2\pi}{a}p\right)r_{\parallel}},\label{eq:b3}\end{equation}
respectively. In Eq. (\ref{eq:b2}), integration over $r_{\parallel}$
immediately yields\begin{equation}
I_{1}^{\left(\alpha\right)}=2\pi\sum_{m=-\infty}^{\infty}c_{m}^{(\alpha)\star}\left(\mathbf{k};n\right)\delta\left(k_{\parallel}-\frac{\omega}{\left|\mathbf{v}\right|}-\frac{2\pi}{a}m\right).\label{eq:b4}\end{equation}
Integral in (\ref{eq:b3}) is equal to $l$, if $k_{\parallel}-\frac{\omega}{\left|\mathbf{v}\right|}-\frac{2\pi}{a}p=0$.
Overwise it results in \begin{equation}
\frac{\sin\left(\left(k_{\parallel}-\frac{\omega}{\left|\mathbf{v}\right|}-\frac{2\pi}{a}p\right)\left(l/2\right)\right)}{\left(k_{\parallel}-\frac{\omega}{\left|\mathbf{v}\right|}-\frac{2\pi}{a}p\right)}.\label{eq:b5}\end{equation}
Then, in the limit $l\rightarrow\infty$, $I_{2}^{\left(\alpha\right)}$
vanishes for $k_{\parallel}-\frac{\omega}{\left|\mathbf{v}\right|}-\frac{2\pi}{a}p\neq0$,
while is equal to $2\pi\sum_{p=-\infty}^{\infty}c_{p}^{(\alpha)}\left(\mathbf{k};n\right)$
for $k_{\parallel}-\frac{\omega}{\left|\mathbf{v}\right|}-\frac{2\pi}{a}p=0$.
Finally, using the function \begin{equation}
\widetilde{\delta}\left(x\right)=\left\{ \begin{array}{c}
1,\,\,\, x=0\\
0,\,\,\, x\neq0\end{array}\right.,\label{eq:b6}\end{equation}
relation (\ref{eq:integrals}) can be written in the form\begin{eqnarray}
I^{\left(\alpha\right)} & =2\pi & q^{2}\sum_{m=-\infty}^{\infty}\sum_{p=-\infty}^{\infty}c_{m}^{(\alpha)\star}\left(\mathbf{k};n\right)c_{p}^{(\alpha)}\left(\mathbf{k};n\right)\times\label{eq:b7}\\
 &  & \delta\left(k_{\parallel}-\frac{\omega}{\left|\mathbf{v}\right|}-\frac{2\pi}{a}m\right)\widetilde{\delta}\left(k_{\parallel}-\frac{\omega}{\left|\mathbf{v}\right|}-\frac{2\pi}{a}p\right),\nonumber \end{eqnarray}
which is non-zero only for $m=p$ yielding relation (\ref{eq:integrals_final}).


\begin{thebibliography}{33}
\bibitem{che34}P. A. Cherenkov, Dokl. Acad. Nauk SSSR \textbf{2},
457 (1934).

\bibitem{jel58}J. V. Jelley, \emph{Cherenkov Radiation and its Applications}
(Pergamon, New York, 1958).

\bibitem{fer40}E. Fermi, Phys. Rev. \textbf{57}, 485 (1940).

\bibitem{akm99}G. N. Afanasiev, V. G. Kartavenko, and E. N. Magar,
Physica \textbf{B269}, 95 (1999).

\bibitem{carb01}I. Carusotto, M. Artoni, G. C. La Rocca, and F. Bassani,
Phys. Rev. Lett. \textbf{87}, 064801 (2001).

\bibitem{acrb03}M. Artoni, I. Carusotto, G. C. La Rocca, and F. Bassani,
Phys. Rev. E \textbf{67}, 046609 (2003).

\bibitem{ft37}I. Frank, I. Tamm, Dokl. Acad. Nauk SSSR \textbf{14},
107 (1937).

\bibitem{jmw95}J. D. Joannopoulos, R. D. Meade, and J. N. Winn, \emph{Photonic
Crystals: Molding the Flow of Light} (Princeton University Press,
Princeton, NJ, 1995).

\bibitem{sak01}K. Sakoda, \emph{Optical Properties of Photonic Crystals}
(Springer, Berlin, 2001).

\bibitem{byk72}V. P. Bykov, Sov. Phys. JETP \textbf{35}, 269 (1972).

\bibitem{yab87}E. Yablonovitch, Phys. Rev. Lett. \textbf{58}, 2059
(1987).

\bibitem{jw90}S. John and J. Wang, Phys. Rev. Lett. \textbf{64},
2418 (1990).

\bibitem{rus86}P. St. J. Russell, Appl. Phys. B: Photophys. Laser
Chem. \textbf{39}, 231 (1986).

\bibitem{zen87}R. Zengerle, J. Mod. Opt. \textbf{34}, 1589 (1987).

\bibitem{kos98}H. Kosaka, T. Kawashima, A. Tomita, M. Notomi, T.
Tamamura, T. Sato, and S. Kawakami, Phys. Rev. B \textbf{58}, R10096
(1998). 

\bibitem{kos99}H. Kosaka, T. Kawashima, A. Tomita, M. Notomi, T.
Tamamura, T. Sato, and S. Kawakami, Appl. Phys. Lett. \textbf{74},
1212 (1999).

\bibitem{ep96}P. Etchegoin and R. T. Phillips, Phys. Rev. B \textbf{53},
12674 (1996).

\bibitem{cst01}D. N. Chigrin and C. M. Sotomayor Torres Opt. Spectrosc.
\textbf{91}, 484 (2001).

\bibitem{chi04}D. N. Chigrin, Phys. Rev. E \textbf{70}, 056611 (2004).

\bibitem{bel86}V. A. Belyakov, Nuc. Inst. Meth. Phys. Res. A \textbf{248},
20 (1986).

\bibitem{abajo03b}F. J. Garcia de Abajo and L. A. Blanco, Phys. Rev.
B \textbf{67}, 125108 (2003).

\bibitem{oo04a}T. Ochiai and K. Ohtaka, Phys. Rev. B \textbf{69},
125106 (2004)

\bibitem{oo04b}T. Ochiai and K. Ohtaka, Phys. Rev. B \textbf{69},
125107 (2004)

\bibitem{yam04}K. Yamamoto, R. Sakakibara, S. Yano, Y. Segawa, Y.
Shibata, K. Ishi, T. Ohsaka, T. Hara, Y. Kondo, H. Miyazaki, F. Hinode,
T. Matsuyama, S. Yamaguchi, and K. Ohtaka, Phys. Rev. E \textbf{69},
045601(R) (2004).

\bibitem{oo06}T. Ochiai and K. Ohtaka, Opt. Express \textbf{14},
7378 (2006).

\bibitem{abajo03}F. J. Garcia de Abajo, N. Zabala,
A. Rivacoba, A.G. Pattantyus-Abraham,  M. O.Wolf, and P.M. Echenique, Phys. Rev. Lett. \textbf{91},
143902 (2003). 

\bibitem{abajo03a}F. J. Garcia de Abajo, A. Rivacoba, N. Zabala, and
P. M. Echenique, Phey Rev. B \textbf{68}, 205105 (2003).

\bibitem{lijj03}C. Luo, M. Ibanescu, S. G. Johnson, and J. D. Joannopoulos,
Science \textbf{299}, 368 (2003).

\bibitem{fer05}V. G. Baryshevsky, I. D. Feranchuk, A. P. Ulyanenkov, \emph{Parametric X-ray radiation
from the electrons in crystals: Theory, Experiment and Applications} (Springer, Heidelberg, 2005).

\bibitem{dow92}J. P. Dowling and C. M. Bowden, Phys. Rev. A \textbf{46},
612 (1992).

\bibitem{jac75}J. D. Jackson, \emph{Classical Electrodynamics} (Wiley,
New York, 1975).

\bibitem{rom03}S. G. Romanov, D. N. Chigrin, V. G. Solovyev, T. Maka,
N. Gaponik, A. Eychmuller, A. L. Rogach, C. M. Sotomayor Torres, Phys.
Stat. Sol. (a) \textbf{197}, 662 (2003).

\bibitem{jj01}S. G. Johnson, and J. D. Joannopoulos, Opt. Express
\textbf{8}, 173-190, (2001).

\bibitem{taf95}A. Taflove, \emph{Computational Electrodynamics: The
Finite-Difference Time-Domain Method} (Artech House, Norwood, 1995)

\bibitem{ber94}J. P. Berenger, J. Comput. Phys. \textbf{114}, 185
(1994).
\end{thebibliography}
\end{document}